# A possible solution of $B \to \pi\pi$ puzzle in $Z'$ model

S. Biswas[1], P. Nayek and S. Sahoo[2]

Department of Physics
National Institute of Technology, Durgapur-713209, West Bengal, India

[1]E-mail: getswagata92@gmail.com, [2]E-mail: sukadevsahoo@yahoo.com

## Abstract

The discrepancies among the measurements of branching ratios and CP asymmetries of $B \to \pi\pi$ decays such as large direct CP asymmetry for $B^0 \to \pi^+\pi^-$ mode and large branching ratio for $B^0 \to \pi^0\pi^0$ mode originate the $B \to \pi\pi$ puzzle. According to diagrammatic approach of this $b \to d$ transition the small ratio of color-suppressed amplitude (*C*) to color-allowed one (*T*) contradicts between the standard model (SM) approach and the experimental values. To make out the ambiguity we need to scrutinize the decays with different topological amplitudes. In this paper, the decays are studied in the SM by taking different values of $C/T$ as constraints. We find that larger ratio of $C/T$ is explained successfully in the SM but the lower ratio is not for which new physics (NP) is needed. The NP contribution can be included in $B \to \pi\pi$ decays at tree level by $Z'$ boson. Here, we find that $Z'$ model can explain the puzzle by providing a good solution for lower ratio of $C/T$.

## 1. Introduction

Decays of B-mesons [1-22] is one of the most promising research areas to obtain information about the flavour structure of the SM as well as to get a track of new physics (NP) path. In recent years, B-meson decays play an important role for theorists as well as experimentalists to study some disagreements between the SM predictions and experimental observations. Different pioneering measurements of B-factories with BaBar and Belle experiments as well as Tevatron provide latest experimental state of B-decays governed by the dedicated B-decays experiment of the Large Hadron Collider LHCb [3, 8-14]. There are three decays of $B \to \pi\pi$ system: $B^+ \to \pi^+\pi^0, B_d^0 \to \pi^+\pi^-$ and $B_d^0 \to \pi^0\pi^0$. Different observables of above three decays have been measured: three branching ratios, three direct CP asymmetries $A_{CP}$ and one mixing-induced CP asymmetry $S_{CP}$. The recent experimental data of branching ratio for $B_d^0 \to \pi^0\pi^0$ is not in agreement with the SM prediction and the value of direct CP asymmetry for $B_d^0 \to \pi^+\pi^-$ is larger than the theoretical value, this is known as "$B \to \pi\pi$ puzzle" [1-3, 15-23]. In Ref. [22], it is observed that the inconsistency with the SM predictions can be explained by performing a full fit with the obtained experimental data as constraints. But every time the fit provides poor value. As we know the quality of fits can be improved by enhancing the number of constraints. So the value of CKM phase angle $\gamma$ is added as a constraint and it recovers the picture of SM explanation up to some extent. Still there are some deficiencies in the fit. Then we have tried for various values of ratio of colour-suppressed amplitude ($C$) and colour-allowed one ($T$). Observing the fit results we can argue that the SM is good in explaining the $B \to \pi\pi$ puzzle but in a particular condition.

This suggests the presence of NP and allows us to test the NP condition for which the puzzle can be fruitfully explained [22, 24-26]. In this work, we have used the updated experimental results to perform the fitting. After getting the fitted results we identify a specific region where SM is able to explain the puzzle and in other region the SM results become terrible. So certainly NP is required. From recent studies [27, 28] it is observed that many NP models involve the exchange of a $Z'$ boson or a leptoquark [29, 30]. NP models introduce additional couplings to new heavy mediators at both tree and loop level and these couplings could modify the values of branching ratios and/or CP asymmetries with respect to their SM values. Here, we consider a $Z'$ model and study the effect of $Z'$-mediated FCNCs on $B \to \pi\pi$ decays.

The NP is allowed to contribute at tree level by $Z'$-mediated flavour changing $B \to \pi\pi$ decays where $Z'$ boson couples to the flavour-changing part $\bar{d}b$ as well as to $\bar{u}u$ and $\bar{d}d$. Including several constraints the fit results are improved and the applied constraints may provide precise couplings that can solve the well-known $B \to \pi\pi$ puzzle. LHCb, ATLAS and CMS experiments at the LHC assemble a huge amount of experimental data of many observables on various rare b-hadron decays. Recently some experimentally observed parameters which show inconsistency from the SM are: (i) observation of lepton flavour universality (LFU) violation in $R_K$ [31], (ii) $R_K^*$ [32], (iii) branching ratio and angular distribution of $B_s^0 \to \varphi\mu^+\mu^-$ [33, 34]. These disagreements of measurements are successfully explained by the $Z'$ models. This type of $Z'$ model is proposed here to establish a potentially furnish explanation of the puzzle.

This paper is structured as follows. In Sec. 2, we discuss the amplitude of $B \to \pi\pi$ decays in terms of topological amplitudes which contain color-suppressed as well as color-allowed ones for both tree and electroweak penguin amplitudes. In Sec. 3, we recapitulate the naive $B \to \pi\pi$ puzzle. In Sec. 4, we briefly discuss the statistics that we have used to present an interesting view of the puzzle. Several fits are done for different values of $C/T$ in Sec. 5. We have summarized the results of SM fits in this section. The NP explanation for the puzzle in $Z'$ model with several NP fits is presented in Sec. 6. The results of NP fits are also summarized in this section. We present our conclusions in Sec. 7.

## 2. Amplitude Structures and Observables of $B \to \pi\pi$ decays

In the SM, the $B \to \pi\pi$ decays involve $b \to d\bar{q}q$ (q = u, d) transitions through exchange of $W$-boson. In the $B \to \pi\pi$ decays [35], the $B$ meson is heavy, sitting at rest. It decays into two light mesons with large momenta. Therefore the light mesons are moving very fast in the rest frame of $B$ meson. The topological amplitudes provide a parameterization for nonleptonic B-meson decay processes. To present the $B \to \pi\pi$ decays in terms of topological amplitudes, we include three types of amplitudes – a "tree" contribution $t$, a "color-suppressed" contribution $c$ and a "penguin" contribution $p$ (here we have neglected annihilation, exchange and penguin-annihilation contributions as it is very small in the SM). These amplitudes contain both leading order and electroweak penguin contributions. Taking the gluonic amplitude smaller than QCD penguin contribution, the amplitudes can be parameterized as [2]:

$$t = T + P_{EW}^C,$$
$$c = C + P_{EW},$$
$$p = P - P_{tu} - \frac{1}{3}P_{EW}^C,$$

where $T$ and $C$ are color-allowed and color-suppressed tree amplitudes and $P_{EW}$ and $P_{EW}^C$ are color-allowed and color-suppressed electroweak penguin amplitudes respectively. Here we consider that the weak factors of the term $P_{tu}$ are same as the tree amplitudes $T$ and $C$, so the term can be allowed to involve within the tree amplitudes. This consideration enhances fit quality which is used in this work. The amplitudes can be written as:

$$t = T + P_{EW}^C,$$
$$c = C + P_{EW},$$
$$p = P - P_{EW}^C, \tag{1}$$

The diagrammatic approach [4, 5] of $B \to \pi\pi$ decay amplitudes can be represented as:

$$A^{+0} = -\frac{1}{\sqrt{2}}(t + c)$$
$$A^{\pm} = -(t + p)$$
$$A^{00} = -\frac{1}{\sqrt{2}}(c - p). \tag{2}$$

Considering the SM approach, we have used $SU(3)$ flavour symmetry to relate $P_{EW}$ and $P_{EW}^C$ to $T$ and $C$ [36-38] termed as EWP-tree relations:

$$P_{EW} = \frac{3}{2}\frac{C_9}{C_1}RT,$$
$$P_{EW}^C = \frac{3}{2}\frac{C_9}{C_1}RC, \tag{3}$$

where, $R = \left|\frac{V_{tb}^* V_{td}}{V_{ub}^* V_{ud}}\right| = 2.19$ [39] and $C_9$ and $C_1$ are Wilson coefficients.

Fixing the phase [40-43] we consider,

$$T = |T|e^{i(\delta_T + \gamma)},$$
$$C = |C|e^{i(\delta_T + \delta_C + \gamma)},$$
$$P = -|P|. \tag{4}$$

Here $\delta_T$, $\delta_C$ are the strong phases and $\gamma$ is the CKM weak phase angle. The negative sign comes for the weak phase $\pi$ associated with $P$. Taking the above considerations into account branching ratios (Br) and CP asymmetries ($A_{CP}$, $S_{CP}$) are represented essentially in terms of diagrams and phases. To structure the branching ratios and CP asymmetries we have used the following expressions.

The partial decay width for $B \to \pi\pi$ decay [2] can be expressed as,

$$\Gamma(B \to \pi\pi) = \frac{p_c}{8\pi m_B^2} |A(B \to \pi\pi)|^2, \tag{5}$$

where $p_c$ is the momentum of $\pi$ meson in the rest frame of B meson, $m_B$ is the mass of B meson. To relate partial decay widths with branching ratios, we use the life times of B mesons [44].

The direct CP asymmetry [45] expression is given as

$$A_{CP} = \frac{|\bar{A}|^2 - |A|^2}{|\bar{A}|^2 + |A|^2}, \tag{6}$$

and the mixing-induced CP asymmetry expression is given as

$$S_{CP} = \frac{2 Im(\bar{A} A^*)}{|A|^2 + |\bar{A}|^2}. \tag{7}$$

Another ingredient that is required for the fitting is the constraints. Here, we have recorded the recent experimental results of the observables which are defined above in Table1 [1, 23].

**Table-1: Experimental data of branching ratios and CP asymmetries**

| Mode | Branching ratio ($10^{-6}$) | $A_{CP}$ | $S_{CP}$ |
| --- | --- | --- | --- |
| $B^+ \to \pi^+ \pi^0$ | $5.5 \pm 0.4$ | $0.03 \pm 0.04$ | — |
| $B_d^0 \to \pi^+ \pi^-$ | $5.12 \pm 0.19$ | $0.31 \pm 0.05$ | $-0.67 \pm 0.06$ |
| $B_d^0 \to \pi^0 \pi^0$ | $1.9 \pm 0.5$ | $0.43 \pm 0.24$ | — |

It should be noted that the mixing-induced CP asymmetry in $B_d^0 \to \pi^+ \pi^-$ depends on the weak phase $\alpha$. The above observables constrain the $B \to \pi\pi$ decays and in addition to these we have also considered the weak phase angles as constraints to improve the fit quality. The independent measurements of the CKM phases are recorded as below [46]:

$$\alpha = (88.8 \pm 2.3)° \text{ and } \gamma = (72.2^{+6.8}_{-7.3})°. \tag{8}$$

## 3. Naive Explanation of $B \to \pi\pi$ puzzle

The main discrepancies observed in $B \to \pi\pi$ decays are as follows [1, 3, 23]:

i. The direct CP asymmetry for $B_d^0 \to \pi^+ \pi^-$ is very large in comparison to the SM value.

ii. The branching ratio has larger value for $B_d^0 \to \pi^0 \pi^0$ than theoretical expectations.

Now we want to explain the puzzle in terms of topological amplitudes. According to the diagrammatic approach [4, 5] the amplitudes are structured using eq. (1) and eq. (2). These diagrammatic amplitudes are expected to be hierarchical in size which is $C$ much less than $T$. On the other hand, the experimental data of branching ratio and CP asymmetry of $B_d^0 \to \pi^0\pi^0$ and $B_d^0 \to \pi^+\pi^-$ respectively demands the large ratio of $C$ to $T$. It is in contradiction to the hierarchy. This is the Naive $B \to \pi\pi$ puzzle which is the centre of interest for decades. The main objective of our paper is to modify the naive explanation by using $Z'$ model and to get a possible solution of the puzzle.

## 4. Statistical analysis

To solve the puzzle, we need to change its usual pattern of representation. In this regard, we study the decays in a statistical method by performing $\chi^2$ fitting including some theoretical inputs. The value of $\chi^2$ is used to find the deviation of experimental values from the expected values. It can be defined as [7, 47-49]:

$$\chi^2 = \sum_i \frac{\left(f_i^{th} - f_i^{exp}\right)^2}{(\Delta f_i)^2}, \qquad (9)$$

where $f_i^{th}$, $f_i^{exp}$ and $\Delta f_i$ represent theoretical expressions of various observables, experimental values and the experimental errors of corresponding observables respectively. By minimising this $\chi^2$ expression we obtain several best fit values. We use the tMinuit package (i.e. Ifit) of ROOT software for fitting. In order to construct a good fit we need to obey the basic rules of $\chi^2$ distribution. So we have to calculate the probability distribution for $\chi^2$ which is defined as [2, 7]:

$$P(\chi^2) = \int_{\chi^2}^{\infty} \frac{1}{2^{n/2}\Gamma(n/2)} (\chi^2)^{\left(\frac{n}{2}-1\right)} e^{-\chi^2/2} d(\chi^2) . \qquad (10)$$

This probability depends on the parameter $'n'$, i.e., degrees of freedom. To avoid statistical fluctuations we define $\chi^2_{min}/n \approx 1$ as acceptable fit condition [43-45] and the probability value is also preferred as 50% (0.5). Here we find that if we increase the numbers of degrees of freedom our fit becomes more fruitful. But for smaller values of degrees of freedom it is difficult to judge the fit quality by using $\chi^2_{min}/n \approx 1$ condition only. In that case the probability values (p-value) will help to determine the goodness of fit results.

## 5. The SM fits

In order to calculate the value of $\chi^2$ as given in Eq. (9), we need $f_i^{th}$. For this, we use Eq. (1) and Eq. (2) in terms of diagrams and phases. So we have 7 observables in terms of amplitudes. Hence, we get the $\chi^2$ expression as a function of diagrams and phases. By minimising this $\chi^2$ expression we get different sets of best fit values for several theoretical inputs. Here, we have 3 branching ratios, 3 CP asymmetries, 1 mixing-induced CP asymmetry and CKM phase angles $\alpha$ and $\gamma$ as known parameters. And the three magnitudes ($C, P, T$) and two strong phases as unknown parameters. Thus, we have more known parameters than unknown parameters, so it is possible to fit the parameters.

Here, we have noted that according to our approach we have got the amplitudes and their topological contributions as given in eq. (1) and eq. (2). Using these expressions the direct CP asymmetry value for $B^+ \to \pi^+\pi^0$ becomes zero in SM. Thus the number of constraints becomes 8 and number of unknowns is 5 (the topological amplitudes and the strong phases), we get degrees of freedom 3. Putting these constraints several best fit values are recorded in the Table-2. If we study the best fit values of Table-2 we can say that the ratio of the contribution of color-suppressed tree amplitude to the color-allowed one is larger which contradicts the diagram. On the other hand, the obtained p-value is also not acceptable.

**Table-2: Best fit values with no theoretical input and constraints: data taken from Table-1, $\alpha$ and $\gamma$ values.**

| $\dfrac{\chi^2_{min}}{n} = \dfrac{0.99}{3}$ p-value $= 0.8037$ ||
|---|---|
| Parameter | Best-fit value |
| $C$ | $0.7 \pm 0.11$ |
| $P$ | $52.8 \pm 4.2$ |
| $T$ | $1.36 \pm 0.123$ |
| $\delta_C$ | $-123.73 \pm 0.052$ |
| $\delta_T$ | $22.559 \pm 0.0627$ |

Here, we need to include some theoretical inputs [50-54]. We have considered the ratio $C/T$ as a key term to make out the $B \to \pi\pi$ puzzle originally. Therefore different values of $C/T$ are included as theoretical inputs along with the constraints in our fits. Gradually we have decreased the ratio value and performed several fits. At first, for the ratio $C/T = 0.7$ we want to scrutinize the topological contributions on the puzzle and the fit results are recorded in the Table-3.

**Table-3: Best fit values with theoretical input $C/T = 0.7$ and constraints: data taken from Table-1, $\alpha$ and $\gamma$ values.**

| $\frac{C}{T} = 0.7$: $\frac{\chi^2_{min}}{n} = \frac{1.99}{4}$ p-value = 0.754 | |
|---|---|
| Parameter | Best-fit value |
| $P$ | $28.23 \pm 1.457$ |
| $T$ | $1.069 \pm 0.073$ |
| $\delta_C$ | $-121.879 \pm 0.0696$ |
| $\delta_T$ | $31.969 \pm 0.042$ |

Now for the ratio $C/T = 0.5$ we find that the situation is much better than the previous one and the best fit values are also good. The fit results are recorded in the Table-4.

**Table-4: Best fit values with theoretical input $C/T = 0.5$ and constraints: data taken from Table-1, $\alpha$ and $\gamma$ values.**

| $\frac{C}{T} = 0.5$: $\frac{\chi^2_{min}}{n} = \frac{2.99}{4}$ p-value = 0.5845 | |
|---|---|
| Parameter | Best-fit value |
| $P$ | $47.59 \pm 1.198$ |
| $T$ | $1.22 \pm 0.057$ |
| $\delta_C$ | $-122.79 \pm 0.046$ |
| $\delta_T$ | $22.603 \pm 0.057$ |

**Table-5: Best fit values with $C/T = 0.2$**

| $C/T = 0.2$: $\chi^2_{min}/n = 67.34/4$ | |
|---|---|
| p-value = $(8.25 \times 10^{-14})$ | |
| Parameter | Best-fit value |
| $P$ | $30.12 \pm 0.647$ |
| $T$ | $0.746 \pm 0.0356$ |
| $\delta_C$ | $-123.45 \pm 0.1498$ |
| $\delta_T$ | $22.603 \pm 0.057$ |

**Table-6: Best fit values with $C/T = 0.1$**

| $C/T = 0.1$: $\chi^2_{min}/n = 107.52/4$ | |
|---|---|
| p-value $\approx 0$ | |
| Parameter | Best-fit value |
| $P$ | $27.23 \pm 0.556$ |
| $T$ | $0.68 \pm 0.028$ |
| $\delta_C$ | $-123.67 \pm 0.2055$ |
| $\delta_T$ | $22.92 \pm 0.0914$ |

We are continuing the decrement of the ratio and we have fitted with theoretical input $C/T = 0.2$. We get a very poor fit. From the Table-5, we have got $\chi^2_{min}/n = 67.34/4$ and the corresponding p-value is $(8.25 \times 10^{-14})$ which is very much far from the acceptable value. Again if we decrease the ratio slightly and fit the $\chi^2$ expression taking $C/T = 0.1$ as theoretical input and experimental values from table as constraints then we find that the fitting is much poorer than the previous fit. The fit results for both $C/T = 0.2$ and $C/T = 0.1$ are recorded in the Table-5 and Table-6 respectively. Here constraints are the data taken from Table-1 and the values of $\alpha$ and $\gamma$.

Observing all the fit results we can conclude that there is no puzzle in the region where $C/T$ has higher values (0.7 and 0.5). Though the p-value (75.4%, from Table-3) for $C/T = 0.7$ is not much good but that is not unphysical and for $C/T = 0.5$ we get p-value 58.45% (from Table-4) which signifies a very good fitting. So we see that the main discrepancy is found in the lower value of the ratio (0.2 and 0.1) obtaining the unphysical p-values. As the SM cannot explain the data in its theoretically-allowed range then we have to include the NP. We have to investigate how the NP can explain the data and in which condition too. In next section, we have started to examine the puzzle with the aspects of NP.

## 6. NP in $B \rightarrow \pi\pi$ puzzle

Let us consider that the NP is included to the amplitudes of $B \rightarrow \pi\pi$ decays through the form $\bar{d}\Gamma_i b \bar{q}\Gamma_j q$ where, $\Gamma_i$ and $\Gamma_j$ represent Lorentz structures and here the color indices are suppressed. The contribution of NP can be structured in the matrix form as $<\pi\pi|\bar{d}\Gamma_i b \bar{q}\Gamma_j q|B>$. Here, we have to note that each matrix term has different strong phases as well as weak phases. From the SM fits, we can conclude that the strong phase associated with $T$, i.e. $\delta_T$ is small (which is originated due to QCD scattering). The primary condition to perform fit is to have more constraints than unknowns. So for simplification we have neglected NP strong phases and considered the NP weak phases only. The above NP operator contributed in two ways [22, 55]:

i. $\overline{d_\alpha}\Gamma_i b_\beta \overline{q_\beta}\Gamma_j q_\alpha$ contains suppression factors with color octet currents. For this contribution the final state must be $\bar{d}q$ meson. The NP amplitude term associated with this contribution is $A^{c,q}e^{i\varphi_q^c}$.

ii. $\overline{d_\alpha}\Gamma_i b_\alpha \overline{q_\beta}\Gamma_j q_\beta$ does not contain any color suppression factor. For this contribution the final state must be $\bar{q}q$ meson. And the NP amplitude term associated with this contribution is $A^q e^{i\varphi_q}$.

In the NP amplitudes, $\varphi_q$ and $\varphi_q^c$ are the NP weak phases.

According to the quark structures of the final state mesons, we have structured the NP amplitudes which are given as,

$$A^{+0} = T + P_{EW}^C + C + P_{EW} + (-A^u e^{i\varphi_u}) + A^d e^{i\varphi_d} + (-A^{c,u}e^{i\varphi_u^c}) + A^{c,d}e^{i\varphi_d^c}$$

$$A^{\pm} = C + P_{EW} + P - \frac{1}{3}P_{EW}^C + (-A^{c,u}e^{i\varphi_u^c})$$

$$A^{00} = C + P_{EW} - P - \frac{1}{3}P_{EW}^C + (-A^u e^{i\varphi_u}) + A^d e^{i\varphi_d} + A^{c,d}e^{i\varphi_d^c} \tag{11}$$

Another set of NP operators can also be defined as below [56]:

$$P_{NP}e^{i\varphi_P} = \frac{1}{3}A^{c,u}e^{i\varphi_u^c} + \frac{2}{3}A^{c,d}e^{i\varphi_d^c}$$

$$P_{EW.NP}^C e^{i\varphi_{EW}^C} = A^{c,u}e^{i\varphi_u^c} - A^{c,d}e^{i\varphi_d^c}$$

$$P_{EW,NP}e^{i\varphi_{EW}} = A^u e^{i\varphi_u} - A^d e^{i\varphi_d} \tag{12}$$

Using the eq. (11) and eq. (12) we finally get the decay amplitudes as,

$$A^{+0} = T + P_{EW}^C + C + P_{EW} - P_{EW,NP}e^{i\varphi_{EW}} - P_{EW.NP}^C e^{i\varphi_{EW}^C}$$

$$A^{\pm} = C + P_{EW} + P - \frac{1}{3}P_{EW}^C - P_{NP}e^{i\varphi_P} - \frac{2}{3}P_{EW.NP}^C e^{i\varphi_{EW}^C}$$

$$A^{00} = C + P_{EW} - P - \frac{1}{3}P_{EW}^C - P_{EW,NP}e^{i\varphi_{EW}} + P_{NP}e^{i\varphi_P} - \frac{1}{3}P_{EW.NP}^C e^{i\varphi_{EW}^C}. \tag{13}$$

Further, the number of NP weak phases ($\varphi_{EW}, \varphi_{EW}^C, \varphi_P$) increase the number of unknowns than the number of constraints. That is why we need to put a new consideration. The four fermion operator corresponding to quark transition $b \to d\bar{q}q$ is proportional to $g_L^{bd}g_{L(R)}^{qq}$. $g_{L(R)}^{qq}$ must be real and it arises due to self-conjugation by the current $\bar{q}\gamma^\mu P_{L(R)}q$. But the weak phase is associated with the $g_L^{bd}$ term (which must be complex) and this weak phase is the source of CP violation in the NP operators where the puzzle is actually concentrated. So we can consider that the all weak phases for all operators are equal and there is only one NP weak phase. Therefore, we have considered $\varphi_d = \varphi_u = \varphi_d^c = \varphi_u^c = \varphi$ and $\varphi_{EW} = \varphi_{EW}^C = \varphi_P = \varphi$ for the two sets of NP decay amplitudes.

## 6.1 Role of $Z'$ model

The quark transition for the $B \to \pi\pi$ decays is $b \to d\bar{q}q$ (where $q = u, d$). In $Z'$ model this decay has been tried to explain via tree-level exchange of $Z'$ boson [57, 58]. Here $Z'$ couples to $b\bar{d}$ and $\bar{q}q$. The flavour-changing coupling $b\bar{d}Z'$ can be considered as a trivial part as NP weak phase [59] dominates which in turn reduces the contribution $g_L^{bd}$. So the coupling $\bar{q}qZ'$ is important in this discussion. Now we investigate how the $Z'$ boson couples to $\bar{q}q$ and for which condition we will get acceptable fit.

Here, in the $Z'$ model, we consider the $SU(2)_L$ symmetry for which we get $g_L^{dd} = g_L^{uu}$. The scenarios we use in our consideration are:

i) All the three NP operators are nonzero with theoretical input $C/T = 0.2$.
ii) If it is required we can increase the color contribution by putting $P_{EWNP}^C < P_{EWNP}$ as an additional theoretical input.
iii) If $Z'$ boson couples vectorially to quark structure $\bar{d}d$ and $\bar{u}u$ then we consider the theoretical input as $P_{EWNP}^C = P_{EWNP} = 0$ and $P_{NP} \neq 0$.
iv) And the last consideration is $P_{NP} = 0$ when the right-handed couplings are nonzero.

## 6.2 The NP fits

Here, we begin with the 1$^{st}$ consideration in which all the three NP operators are nonzero. The p-value is 57.77% and so we can say that the fitting is good for $C/T = 0.2$ and the fit results are given in Table-7. From Table-7, we find that the ratio $P_{EWNP}^C / P_{EWNP}$ is 0.949 which implies that the color allowed electroweak penguin amplitude and color suppressed electroweak penguin amplitude contributes almost same amount. But our main objective is to inspect the puzzle using NP with color contribution. So here we need to increase the color contribution by inputting $P_{EWNP}^C / P_{EWNP} = 0.3$ as theoretical input. The fit results for this consideration are given in the Table-8. From this table, it is clear that we get the p-value equal to 48.22% which is quite good. Therefore we can say that the fitting is acceptable. The color contribution is successfully explained by the NP with good fit values. Here, it can be noted that we have increased the fitting quality by increasing the number of degrees of freedom and by specifying the investigation area with the theoretical input $P_{EWNP}^C / P_{EWNP} = 0.3$.

Let us consider the 3$^{rd}$ scenario. Here, we consider that the $Z'$ boson couples to the quark structure $\bar{d}d$ and $\bar{u}u$ vectorially. This implies that we consider all the right handed and left handed couplings for both $\bar{u}u$ and $\bar{d}d$ which are equal, i.e., $g_L^{dd} = g_R^{dd} = g_L^{uu} = g_R^{uu}$. For this case the NP amplitudes in eq. (11) become as $A^{c,u} = A^{c,d}$ and $A^u = A^d$. It provides another set of NP amplitudes in eq. (12) as $P_{EWNP}^C = P_{EWNP} = 0$. On the other hand, if we consider negligible right handed couplings i.e., $g_R^{dd} = g_R^{uu} = 0$ then also we will get same conditions. In this scenario the only nonzero parameter is $P_{NP}$ and the fit results are recorded in the Table-9.

**Table-7: Best fit values with theoretical input $C/T = 0.2$ and constraints are taken from Table-1, $\alpha$ and $\gamma$ values.**

| $\frac{C}{T} = 0.2$: $\frac{\chi^2_{min}}{n} = \frac{0.31}{1}$ p-value = 0.5777 | |
|---|---|
| Parameter | Best-fit value |
| $P$ | $23.91 \pm 3.144$ |
| $T$ | $0.78 \pm 0.0847$ |
| $\delta_C$ | $-123.169 \pm 0.6087$ |
| $\delta_T$ | $24.266 \pm 0.3087$ |
| $P_{EWNP}$ | $16.81 \pm 1.8474$ |
| $P^C_{EWNP}$ | $15.95 \pm 1.8477$ |
| $P_{NP}$ | $9.321 \pm 1.202$ |
| $\varphi$ | $70.53 \pm 0.0796$ |

**Table-8: Best fit values with theoretical inputs $C/T = 0.2$, $P^C_{EWNP}/P_{EWNP} = 0.3$ and constraints: data taken from Table-1, $\alpha$ and $\gamma$ values.**

| $\frac{C}{T} = 0.2, \frac{P^C_{EWNP}}{P_{EWNP}} = 0.3$: $\frac{\chi^2_{min}}{n} = \frac{1.4586}{1}$, p-value = 0.4822 | |
|---|---|
| Parameter | Best-fit value |
| $P$ | $25.576 \pm 2.43$ |
| $T$ | $0.798 \pm 0.062$ |
| $\delta_C$ | $-123.08 \pm 0.473$ |
| $\delta_T$ | $24.17 \pm 0.4$ |
| $P_{EWNP}$ | $0.704 \pm 0.793$ |
| $P_{NP}$ | $15.026 \pm 10.20$ |
| $\varphi$ | $77.0699 \pm 0.197$ |

**Table-9: Best fit values with theoretical inputs $C/T = 0.2$, $P^C_{EWNP} = P_{EWNP} = 0$, $P_{NP} \neq 0$ and constraints: data taken from Table-1, $\alpha$ and $\gamma$ values.**

| $\frac{C}{T} = 0.2$, $P^C_{EWNP} = P_{EWNP} = 0$, $P_{NP} \neq 0$: $\frac{\chi^2_{min}}{n} = \frac{2.39}{3}$, p-value = 0.4955 | |
|---|---|
| Parameter | Best-fit value |
| $P$ | $25.53 \pm 2.0177$ |
| $T$ | $0.775 \pm 0.025$ |
| $\delta_C$ | $-123.11 \pm 0.027$ |
| $\delta_T$ | $24.01 \pm 0.161$ |
| $P_{NP}$ | $19.61 \pm 5.327$ |
| $\varphi$ | $73.97 \pm 0.0347$ |

**Table-10: Best fit values with theoretical inputs $C/T = 0.2$, $P^C_{EWNP}/P_{EWNP} = 0.3$, $P_{NP} = 0$ and constraints: data taken from Table-1, $\alpha$ and $\gamma$ values.**

| $\frac{C}{T} = 0.2$, $\frac{P^C_{EWNP}}{P_{EWNP}} = 0.3$, $P_{NP} = 0$: $\frac{\chi^2_{min}}{n} = \frac{150.831}{3}$, p-value $= -1.77 \times 10^{-16}$ | |
|---|---|
| Parameter | Best-fit value |
| $P$ | $26.37 \pm 1.75$ |
| $T$ | $0.6 \pm 0.16$ |
| $\delta_C$ | $-123.008 \pm 0.40$ |
| $\delta_T$ | $23.816 \pm 0.11$ |
| $P_{EWNP}$ | $26.24 \pm 1.85$ |
| $\varphi$ | $78.25 \pm 0.03$ |

From the fit results of the Table-9 we see that the p-value is 49.55% which implies that the fitting is excellent. We have got the fit values also good. But still now we cannot conclude that $Z'$ boson couples to the left handed quarks only or they couples vectorially with both. To make out the matter what is going on, we need to come to the 4$^{th}$ scenario where we have considered that $g_R^{dd}$ and $g_R^{uu}$ are nonzero. Not only that the couplings also hold the relation $g_R^{uu} = -2g_R^{dd}$ and this provides the condition $P_{NP} = 0$. Using all these thoughts we have performed the fitting and the obtained fit results are recorded in the Table-10. From the fitting results we get $\chi^2_{min}/n = 150.831/3$ and p-value is negative which is totally unphysical. In spite of using NP the fitting is very much poor. So it can be said that there are no couplings between the right handed quarks and $Z'$ boson.

Now we summarize the results of our NP fits. The smaller ratio of the key term $C/T$ is successfully explained after introducing the NP terms in the decay amplitudes and the fitting is produced the p-value as 57% (Table-7). Then we have increased the color contribution in electroweak penguin amplitudes and obtained a good fit with p-value 48.22% (Table-8). Now we know that the NP (in $Z'$ model) can explain the puzzle very well but we do not know how the $Z'$ boson couples with the quarks. To get the answer we have again done the fittings separately for right handed couplings as well as for left handed couplings. And according to the result of Table-10 we have obtained a very poor fit with unphysical results. So right handed couplings are excluded and the fine result of Table-9 offered a secured conclusion that the $Z'$ boson couples to the left handed quarks only.

## 7. Conclusions

In this paper, we have studied the $B \to \pi\pi$ puzzle in an updated version as well as looked for a NP explanation by doing different fits with different theoretical inputs. The contradiction between the diagrammatic approach and experimental values of $B \to \pi\pi$ decays is expressed by the key term of the puzzle $C/T$. The SM has successfully explained the higher ratio ($C/T = 0.7, 0.5$) and obtained p-values 75.4% and 58.77% respectively. These fits are quite good. But the problem is with the lower ratios and for those cases ($C/T = 0.2, 0.1$) we have obtained unphysical p-values. To explain these cases we have introduced the NP. At tree level the NP is contributed in $B \to \pi\pi$ decays by $Z'$ boson. Here, we have considered that the decay $b \to d\bar{q}q$ is mediated by $Z'$ boson which couples to $\bar{d}b$ as well as $\bar{q}q$ ($q = u, d$). We have found that the $Z'$ model explains the discrepancies of data with diagrams completely. The color contribution on the decays is also fully explained. And the fit result of the Table-9 predicts that the particular $Z'$ model where the $Z'$ couples to the left handed quarks only have more potential to solve this well known $B \to \pi\pi$ puzzle. As the fitting of Table-9 is very fine having p-value 49.55%, we can say that the puzzle can be solved by the model where the $Z'$ couples to the left handed quarks. There are some models [60-63] which are in agreement with our result. Furthermore, our $Z'$ model can be used to solve the recently observed anomalies in B meson sector such as (i) $R_K$, (ii) $R_K^*$, (iii) branching ratio and angular distribution of $B_s^0 \to \varphi\mu^+\mu^-$ and (iv) angular distribution of $B_d^0 \to \rho l^+l^-$ and $B \to \pi l^+l^-$ [64-67].

# Acknowledgement

We thank the reviewers for suggesting valuable improvements of our manuscript. S. Biswas acknowledges NIT Durgapur for providing fellowship for her research. P. Nayek and S. Sahoo would like to thank SERB, DST, Govt. of India for financial support through grant no. EMR/2015/000817. We also thank to Srikanta Tripathy, Institute of Physics, Bhubaneswar, Odisha, India and Sudip Mandal for useful discussions.